\documentclass[conference]{IEEEtran}
\usepackage{amsmath,amssymb}
\usepackage[mathscr]{eucal}
\usepackage{graphics,graphicx,multicol}
\usepackage{color}
\newcommand{\beq}{\begin{eqnarray}}
\newcommand{\eeq}{\end{eqnarray}}
\newcommand{\n}{\nonumber}
\newcommand{\calN}{\mathcal{N}}
\newcommand{\bV}{{\mathbf V}}
\newcommand{\bF}{{\mathbf F}}
\newcommand{\bc}{{\mathbf c}}
\newcommand{\be}{{\mathbf e}}
\newcommand{\bH}{{\mathbf H}}
\newcommand{\bb}{{\mathbf b}}
\newcommand{\bt}{{\mathbf t}}
\newcommand{\ba}{{\mathbf a}}
\newcommand{\bG}{{\mathbf G}}
\newcommand{\bT}{{\mathbf T}}
\newcommand{\bI}{{\mathbf I}}
\newcommand{\bA}{{\mathbf A}}
\newcommand{\bSig}{{\mathbf \Sigma}}
\newcommand{\bbR}{\mathbb{R}}
\newcommand{\bbZ}{\mathbb{Z}}
\DeclareMathOperator{\Exp}{{\mathbb E}}

\newtheorem{thm}{Theorem}

\newtheorem{lem}[thm]{Lemma}

\newtheorem{rem}{Remark}

\IEEEoverridecommandlockouts

\begin{document}
\title{Sum Capacity of $K$ User Gaussian Degraded Interference Channels}
\author{Jubin Jose and Sriram Vishwanath\\
Department of Electrical and Computer Engineering\\
University of Texas at Austin\\
\{jubin, sriram\}@austin.utexas.edu
}
\maketitle

\begin{abstract}
This paper studies a family of genie-MAC (multiple access channel) outer bounds for $K$-user Gaussian interference channels. This family is inspired by existing genie-aided bounding mechanisms, but differs from current approaches in its optimization problem formulation and application. The fundamental idea behind these bounds is to create a group of genie receivers that form multiple access channels that can decode a subset of the original interference channel's messages. The MAC sum capacity of each of the genie receivers provides an outer bound on the sum of rates for this subset. The genie-MAC outer bounds are used to derive new sum-capacity results. In particular, this paper derives sum-capacity in closed-form for the class of $K$-user Gaussian degraded interference channels. The sum-capacity achieving scheme is shown to be a successive interference cancellation scheme. This result generalizes a known result for two-user channels to $K$-user channels.
\end{abstract}


\section{Introduction}
Interest in the interference channel and its fundamental limits stems from the wide range of applications that will benefit from such analysis. However, large gaps exist in our understanding of interference channels. Since the introduction of interference channels \cite{Ahlswede71}, the class of two-transmitter two-receiver interference channels have been studied in great detail. Indeed, a majority of exact capacity results are known only for such two-user interference channels. The most popular achievable strategy is the Han-Kobayashi strategy \cite{Han_Kobayashi}. Special cases of this strategy for Gaussian channels have been shown to be optimal for multiple classes of channels \cite{Sato,Shang,Motahari,Annapureddy}, and to be within one bit in general \cite{ETW}. Genie-aided bounds have played a central role in the successes in this domain \cite{Kramer00}.

For interference channels with more than two users, there is a growing body of work on new achievable rate regions using concepts such as alignment \cite{Cadambe-Jafar,Jafarian-Jose-Vishwanath-09}. However, the literature on outer bounds for these channels is still limited. In the special case of determining the degrees of freedom (DoF) of $K$-user interference channels, effective outer bounds have been developed. In particular, using multiple-access type bounds \cite{Jafar}, the DoF has been shown to be outer bounded by $K/2$. A tighter outer bound has been developed for interference channels with rational channel gains using combinatorial arguments \cite{EO}. However, in the domain of finite signal to noise ratio (SNR) channels, there is limited existing literature on non-trivial outer bounds for this channel.

A majority of the outer bounds for the interference channel can be subdivided into the following inter-related families: The broadcast (BC) type, the MAC type, the ``Z'' interference-channel type, the genie-aided type and the additive-combinatorial type. The first four types have a lot in common, and a good understanding of these techniques for two-user interference channels can be gained from \cite{Kramer00}. The fifth and last type is distinct from the other techniques and has been studied relatively recently \cite{EO}.

In this paper, our first goal is in developing an outer bound that incorporates elements of the MAC type and genie-aided type outer bounds. This is because the MAC type and genie-aided type bounds have proven to be effective in the two-user interference channel literature. In fact, a majority of existing capacity results in the two-user interference channel domain have resulted from the application of these two families of outer bounds \cite{Sato,Annapureddy,Shang,Motahari}. Thus, a next logical step is to better understand their value in the $K$-user Gaussian interference channel setting.

A MAC-type bound provides an outer bound to the original interference channel in terms of an equivalent Gaussian MAC channel. As demonstrated in \cite{Sato}, this bound can be used to determine the capacity of two-user strong interference channels. In addition, it provides a good outer bound on the DoF of $K$-user interference channels \cite{Jafar}. A genie-aided bound provides receivers in the interference channel with one or more ``genies'' (side information), thus transforming the channel into one where the rate region can be characterized in closed form \cite{Kramer00,ETW}. These bounds have proven to be effective for characterizing the sum capacity of very weak interference channels \cite{Annapureddy,Shang,Motahari}.

We develop an outer bound on the capacity region of $K$-user Gaussian interference channels (ICs) by characterizing classes of genie-MAC receivers. Even though, as a concept, MAC-type and genie-aided outer bounds are well-understood, their application and optimization for the case of Gaussian ICs is far from trivial. A $K$-user Gaussian IC has many more parameters than a two-user case (as studied in \cite{Kramer00}) making this optimization an even more involved process. In this paper, our second goal is to demonstrate that the outer bounds developed can prove new capacity results for an important class of $K$-user channels. We introduce new construction-based proof techniques to evaluate the outer bounds for degraded channels, and characterize the sum capacity of this class of channels in closed-form. The class of degraded channels does not belong to previously known classes including ``weak'' and ``strong'' classes. Our result includes the previously known result on the sum capacity of two-user Gaussian degraded ICs \cite{Sato_degraded,Sason}. The earlier proofs do not directly generalize to $K$-user channels. Thus, our new result generalizes the known two-user result to $K$-user Gaussian ICs using the MAC-genie outer bounds developed in this paper.

The rest of this paper is organized as follows: The next section presents the system model. In Section \ref{sec:outer-bound}, we characterize an outer bound on capacity of $K$-user Gaussian interference channel. In Section \ref{sec:applications}, we derive the sum capacity of the class of degraded channels. We conclude with Section \ref{sec:conclude}.

\section{System Model}
\label{sec:system-model}

\begin{figure}[!t]
\centering
\scalebox{0.8}{\input{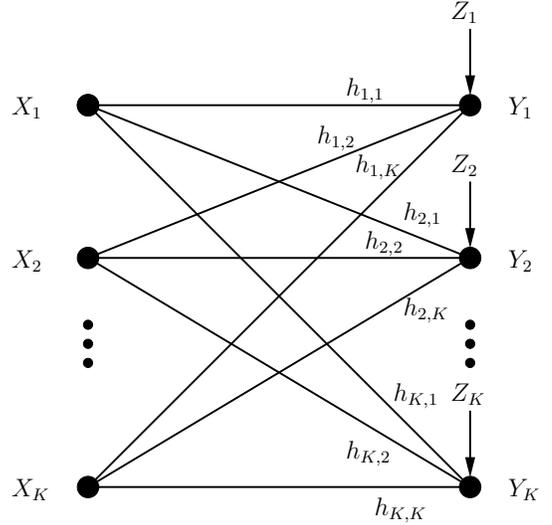}}
\caption{Gaussian $K$-user interference channel}
\label{fig:sysmodel}
\end{figure}

We consider the $K$-user Gaussian interference channel defined as follows: a communication system consisting of $K$ transmitter-receiver pairs labeled $1, 2, \ldots, K$. This channel is shown in Figure \ref{fig:sysmodel}. Each transmitter has independent messages intended for the corresponding receiver. At time $t, t \in \bbZ_+$, the input-output relations that describe the system are:
\beq
\label{eq:gen_int_ch}
Y_i[t] = \sum_{j}h_{i,j}X_j[t] + Z_i[t].
\eeq
Here, $X_j[t]$ is the signal transmitted by the $j$-th transmitter, $h_{i,j}$ is the constant channel gain from $j$-th transmitter to $i$-th receiver, $Z_i[t]$ is the additive white Gaussian noise (AWGN) at $i$-th receiver, and $Y_i[t]$ is the signal received at the $i$-th receiver. For simplicity, we consider real valued signal/gain/noise and suppress the time index $t$ henceforth. The power constraint at the $j$-th transmitter is $\Exp[X_j^2] \le P,$ and the AWGN noise at all receivers have zero mean and variance $N$.

The $K$-user Gaussian interference channel is characterized by $\sqrt{{P}/{N}}\bH$, where $\bH$ is the matrix with $h_{i,j}$ as the entry corresponding to the $i$-th row and the $j$-th column. We use standard information-theoretic definitions for the capacity region and the sum capacity of this channel. Throughout this paper, $C^{\text{IC}}(\sqrt{{P}/{N}}\bH)$ denotes the $K$-dimensional capacity region, $C^{\text{IC}}_{\Sigma}(\sqrt{{P}/{N}}\bH)$ denotes the sum capacity, and  $R_i$ denotes the rate corresponding to the $i$-th transmitter-receiver pair.

\subsection{Notation}
Matrices (and some vectors) are denoted by bold letters. $\bA^*$ denotes the transpose of a matrix $\bA$ and $\bA^+$ denotes its upper triangular portion. $\bA\succ 0$ denotes a symmetric positive-definite matrix. $|\cdot|$ denotes the determinant of a square matrix and the cardinality of a set or vector. $\bI$ denotes the identity matrix. $\Exp[\cdot]$ denotes the expectation operator.

\section{Outer Bound on Capacity Region of $K$-user Interference Channels}
\label{sec:outer-bound}

The main idea behind the outer bound is to adapt the framework in \cite{Kramer00} to the K-user setting. In effect, a {\em genie-MAC} is created to decode a subset of messages in the original interference channel. The capacity region of this genie-MAC channel then forms an outer bound on the rate region of the original channel. This genie-MAC technique is a two-step process. The first step is to find a characterization for the genie-MAC receivers, and the second step is to optimize this characterization to obtain the tightest bound of this class. 

Consider any permutation function $\pi: \{1,2,\ldots,K\} \mapsto \{1,2,\ldots,K\}$, and integers $k$ and $m$ such that $1 \le k \le K$ and $m \ge 1$. Define tuples $S=(\pi(1),\ldots,\pi(k))$ and $S^c=(\pi(k+1),\ldots,\pi(K))$. We use $X_S$ to denote the vector $[X_{S(1)} \,X_{S(2)} \quad X_{S(|S|)}]^*$. Now, consider the multiple-antenna MAC channel that has $X_{S^c}$ as side information at the $m$-antenna receiver and observes the signal
\beq
\label{eq:MAC_ch}
\overline{Y}=\bG X_S + \overline{Z},
\eeq
where $\overline{Z}$ is i.i.d. $\calN(0,\bSig),$ for some $\bG \in \bbR^{m \times k}$. Let $C^{\text{MAC}}(\sqrt{P}\bG,\bSig)$ denote the capacity region of this MAC channel and $C_{\Sigma}^{\text{MAC}}(\sqrt{P}\bG,\bSig)$ denote the sum capacity of this MAC channel. Since the side information is independent of both $X_S$ and $\overline{Z}$, it does not change the capacity region. 

Next, we provide the conditions under which the capacity region of this MAC channel will form an outer bound on $R_S$ of the original interference channel.

\begin{lem}
Consider any $\bT =[\bt_1 \, \bt_2 \quad \bt_k]\in \bbR^{m \times k}$. Let $\bT$, $\bG$ and $\bSig$ be matrices that satisfy the following conditions:
\beq
\left(\bT^*\bG\right)^+ &=& \bH_S^+, \n\\
\bt_i^*\bSig \bt_i &\le& N, \quad \forall 1\le i\le k, \n \\
\bSig &\succ& 0, \n
\eeq
where $\bH_S$ is $|S|\times |S|$ matrix with entry corresponding to the $i$-th row and $j$-th column as $h_{S(i),S(j)}$.
Then,
\[
R_S \in C^{\text{MAC}}(\sqrt{P}\bG,\bSig),
\]
i.e., the capacity region of any MAC channel described by (\ref{eq:MAC_ch}) satisfying the above conditions is an outer bound on the rates $R_S$ for the interference channel described by (\ref{eq:gen_int_ch}).
\end{lem}
\begin{IEEEproof}
We show the following to prove the this lemma. If there exists an achievable strategy for the interference channel described by (\ref{eq:gen_int_ch}) to achieve rates $(R_1,R_2,\cdots,R_K)$, i.e., if $(R_1,R_2,\cdots,R_K) \in C^{\text{IC}}(\sqrt{{P}/{N}}\bH)$, then there exists an achievable strategy for the MAC channel described by (\ref{eq:MAC_ch}) to achieve rates $R_S$, i.e., $R_S\in C^{\text{MAC}}(\sqrt{P}\bG,\bSig)$. In particular, we prove that the MAC channel can obtain statistically identical (or better) signal as (than) $Y_i$ for all $i\in S$.

Let $\bf{D}=\bT^*\bG.$ At the MAC receiver, the signal corresponding to $Y_{S(l)}$ ($1\le l \le k$) is obtained sequentially. Consider any step $l$. Since the messages from transmitters $S(1),S(2),\ldots,S(l-1)$ have been decoded, the receiver can generate signals $X_{S(1)},X_{S(2)},\ldots,X_{S(l-1)}$. In addition, the MAC receiver has signals $X_{S^c}$ as side information. Therefore, the MAC receiver can obtain the signal
\beq
\tilde{Y}_l &=& t^*_l\overline{Y} - \sum_{i=1}^{l-1}d_{l,i}X_{S(i)}+ \sum_{i=1}^{l-1}h_{S(l),S(i)}X_{S(i)} \n\\
&&+ \sum_{i\in S^c}h_{S(l),i}X_{i},\n
\eeq
which can be simplified as
\beq
\tilde{Y}_l&=&\sum_{i=1}^{K}h_{S(l),i}X_{i} + t^*_l\overline{Z}\n.
\eeq
The last step follows from $\left(\bT^*\bG\right)^+ = \bH_S^+$. Since, $t_l^*\bSig t_l \le N$, the MAC receiver can decode the message from transmitter $S(l)$ if the receiver $S(l)$ in the original interference channel can decode the message from transmitter $S(l)$. This completes the proof of lemma.
\end{IEEEproof}

The sum capacity of the MAC channel is given by
\[
C_{\Sigma}^{\text{MAC}}(\sqrt{P}\bG,\bSig)=\frac{1}{2}\log \left(|\bI+P\bSig^{-1}\bG\bG^{*}|\right).
\]
Thus, the minimization problem of interest is
\beq
\label{eq:main_opt_problem}
& f^*(\bH_S, m) & =\inf_{\bG, \bSig,\bT} \frac{1}{2}\log \left(|\bI+P\bSig^{-1}\bG\bG^{*}|\right)\\
&\text{such that}&\left(\bT^*\bG\right)^+ = \bH_S^+, \n\\
&&\bt_i^*\bSig \bt_i \le N, \quad \forall 1\le i\le k, \n \\
&&\bSig \succ 0. \n
\eeq
For $m=|S|$, it is clear that the feasible set is non-empty as $\bG=\bH_S$, $\bSig=N\bI$ and $\bT = \bI$ satisfies all the constraints. We denote this optimization problem with $m=|S|$ by $f^*(\bH_S)$. In the remaining part of this paper, we assume that $m=|S|$.

From the above analysis, we obtain the following theorem that provides an outer bound on the capacity region of the $K$-user Gaussian interference channel. 

\begin{thm}
Consider the interference channel $\bH$ described by (\ref{eq:gen_int_ch}). Then, 
\beq 
C^{\text{IC}}(\sqrt{{P}/{N}}\bH) \subseteq \left\{(R_1,\ldots,R_K): \sum_{i\in S}R_i \le f^*(\bH_S) ,\forall S  \right\}.\n
\eeq
\end{thm}

The above theorem requires the evaluation of the optimization problem given by (\ref{eq:main_opt_problem}). Next, we derive results that simplify this optimization problem. In particular, we show that, any one of the three parameters can be fixed to identity without affecting the optimal value. The next two lemmas prove these results.

\begin{lem}
Consider the following optimization problem that results by choosing $\bSig = \bI$: 
\beq
\label{eq:opt_problem_simp}
&\min_{\bG,\bT} &\frac{1}{2}\log \left(|\bI+P\bG\bG^{*}|\right)\\
&\text{such that}&\left(\bT^*\bG\right)^+ = \bH_S^+, \n\\
&&\bt_i^*\bt_i \le N, \quad \forall 1\le i\le k. \n 
\eeq
\end{lem}
Then, the optimal value of this problem is $f^*(\bH_S)$.

\begin{IEEEproof}
Consider a feasible set of parameters $\bG$, $\bSig = \bA \bA^*$ and $\bT$ for the optimization problem given by (\ref{eq:main_opt_problem}). Let $\hat{\bG}= \bA^{-1}\bG$ and $\hat{\bT}= \bA^{*}\bT$. Now, we have the following:
\beq
\hat{\bT}^*\hat{\bG}=\bT^*\bA\bA^{-1}\bG=\bT^*\bG, \n \\
\hat{\bT}^*\hat{\bT}=\bT^*\bA\bA^{*}\bT=\bT^*\bSig \bT. \n
\eeq
Therefore, $\hat{\bG}$ and $\hat{\bT}$ form a feasible set for the optimization problem given by (\ref{eq:opt_problem_simp}). Furthermore, the objective value remains the same due to the following:
\beq
|\bI+P\hat{\bG}\hat{\bG}^{*}| &=& |\bI+P\bA^{-1}\bG\bG^*\bA^{-1*}| \n \\
&=& |\bI+P\bA^{-1*}\bA^{-1}\bG\bG^*| \n \\
&=& |\bI+P \bSig ^{-1}\bG\bG^*|.\n
\eeq
This completes the proof.
\end{IEEEproof}

\begin{lem}
Consider the optimization problem given by  (\ref{eq:main_opt_problem}). Now, consider 
the two sub-problems resulting from choosing either $\bT = \bI$ or $\bG = \bI$. Then, each of  these sub-problems has optimal value $f^*(\bH_S)$.
\end{lem}

\begin{IEEEproof}
Case-I ($\bT = \bI$): Consider a feasible set of parameters $\hat{\bG}$ and $\hat{\bT}$ for the optimization problem given by (\ref{eq:opt_problem_simp}). Let $\epsilon $ be an arbitrary real number such that $0< \epsilon <1$. Let $\bG= \hat{\bT}^*\hat{\bG}$ and $\bSig= \epsilon N \bI+ (1-\epsilon)\hat{\bT}^*\hat{\bT}$. It is fairly straightforward to check that these parameters are feasible for the sub-problem. Further, the objective value approaches that of the original problem with $\epsilon \rightarrow 0.$

Case-II ($\bG = \bI$): Consider a feasible set of parameters $\hat{\bG}$ and $\hat{\bT}$ for the optimization problem given by (\ref{eq:opt_problem_simp}). Let $\epsilon $ be an arbitrary real number such that $0< \epsilon <1$. Let $\bT= \hat{\bG}^*\hat{\bT}$ and $\bSig= (\epsilon \bI+ \hat{\bG}^*\hat{\bG})^{-1}$. Again, it is fairly straightforward to check that these parameters are feasible for the sub-problem, and the objective value approaches that of the original problem with $\epsilon \rightarrow 0.$
\end{IEEEproof}

Next, we compare this outer bound expression with other techniques in literature. It is fairly simple to see that this bound incorporates receiver cooperation as a special case. In particular, by choosing the matrix $\bG$ to be the same as the channel gains in the original interference channel, the receiver cooperative bound can be obtained. A multiple-access type outer bound as studied in \cite{Sato,Jafar} is also a special case of this bound. A conventional MAC-type bound corresponds to the case when $S$ is a set of the form $\{i,j\}$ and $\bG$ equals the received signal at Receiver $i$ in the original channel. It is perhaps not as straightforward to see that this is, in fact, a genie-aided outer bound. If we were to choose a subset of the rows of the matrix $\bG$ to match those in the original interference channel definition, then the remaining rows of $\bG$ along with $X_{S^c}$ represent a ``vector genie'' provided to enable all messages to be decoded in the system. This bound does not capture all genie-aided bounds in the two-user setting.

Although it captures many existing bounding techniques for the interference channel, the optimization problem in (\ref{eq:opt_problem_simp}) does not necessarily lend itself to a straightforward solution. Furthermore, to evaluate the bound on the sum of a set of rates, we need to consider all possible orderings of tuples $S$ resulting from this set. In the next section, we show that this bound can be evaluated for the class of $K$-user degraded interference channels in closed-form.

\section{Sum Capacity of $K$-user Degraded Interference Channels}
\label{sec:applications}

We study the class of K-user Gaussian degraded interference channels, where degraded is formally defined as the existence of an ordering of the receivers such that the received signals are stochastically degraded in that order. For the Gaussian interference channels, degraded implies unit rank channel matrices. Therefore, all degraded channels can be expressed as $\bH = \ba \bb^*$, where $\ba=[a_1 \: a_2\: 
\ldots \: a_K]^*$ and $\bb=[b_1 \: b_2 \: 
\ldots \: b_K]^*$. Without loss of generality, we assume $a_1^2 \le a_2^2 \le  
\ldots \le a_K^2$, and $P=N=1$. 

\subsection{Achievability}
We consider the successive interference cancellation (SIC) scheme for achievability. Each transmitter uses Gaussian codewords to encode its message. The $i$-th receiver decodes the messages from transmitters $1, 2, \ldots, i$ in this order. Since $i$-th receiver has a (statistically) better received signal than receivers $1,2,\ldots,i-1$, the message at $i$-th transmitter can be encoded at rate
\beq
\label{eq:Rirate}
R_i = \frac{1}{2}\log\left(1+ \frac{a_i^2 b_i^2}{a_i^2\left(\sum_{j=i+1}^{K}b_j^2\right)+1}\right)
\eeq
such that all receivers $i,i+1,\ldots,K$ can decode it with decaying probability of error. Since this is a well-known technique, we do not provide further details. From (\ref{eq:Rirate}), the achievable sum rate using this SIC scheme can be expressed as 
\beq
&&\sum_{i=1}^K R_i = \frac{1}{2} \sum_{i=1}^K \log\left(\frac{a_i^2\left(\sum_{j=i}^{K}b_j^2\right)+1}{a_i^2\left(\sum_{j=i+1}^{K}b_j^2\right)+1}\right), \n \\
&&= \frac{1}{2}  \log\left(\frac{\prod_{i=1}^K\left(a_i^2\left(\sum_{j=i}^{K}b_j^2\right)+1\right)}{\prod_{i=1}^K\left(a_{i-1}^2\left(\sum_{j=i}^{K}b_j^2\right)+1\right)}\right), \n \\
\label{eq:ach_sum_rate}
&&=\frac{1}{2} \sum_{i=1}^K \log\left(1+ \frac{(a_i^2-a_{i-1}^2)\left(\sum_{j=i}^{K}b_j^2\right)}{a_{i-1}^2\left(\sum_{j=i}^{K}b_j^2\right)+1}\right),
\eeq
where $a_0 = 0$ is introduced for notational convenience. 

\subsection{Outer Bound}
The main step is to obtain a matching outer bound on sum rate. We apply the general technique developed in Section \ref{sec:outer-bound} to obtain the outer bound. As discussed before, it is very hard to evaluate these bounds in general, but the degraded structure can be exploited as shown next. 

Consider the optimization problem given by (\ref{eq:opt_problem_simp}) for the tuple $S=(1, 2, \ldots, K)$. Solving this is equivalent to showing the existence of feasible $\bG$ and $\bT$ that evaluates to the right hand side (RHS) of (\ref{eq:ach_sum_rate}). Now, consider the following construction for $\bG$ and $\bT$. Given any $i$ such that $1\le i \le K$, let 
\beq
\label{eq:ci}
c_i = \sqrt{a_i^2 - a_{i-1}^2},
\eeq
and $\bc = [c_1 \: c_2 \: \ldots \: c_K]^*$. We use the following iterative construction to obtain a upper-triangular matrix $\bT$ (lower-triangular $\bT^*$):
\beq
\label{eq:t_construc}
\bt_i = \frac{a_{i-1}}{a_i} \bt_{i-1} + \frac{c_{i}}{a_i} \be_{i},\quad \forall i,
\eeq
where $\bt_{0} = {\mathbf 0}$ and $\be_{i}$ is the unit-vector along $i$-th dimension. The entry corresponding to the $i$-th row and $j$-th column of $\bG$ is chosen as 
\beq
\label{eq:g_construc}
g_{i,j} = c_ib_j d_{i,j},\quad \forall i,j,
\eeq
where $d_{i,j}$ parameters are introduced here for the first time. We fix $d_{i,j} = 1$ for any $i \le j.$ The choice of remaining parameters ($d_{i,j}$ for $i > j$) are discussed later. Irrespective of these remaining parameters, the above construction has the following property.

\begin{lem}
Consider any $\bG$ and $\bT$ given above. Then, it belongs to the feasible set corresponding to the optimization problem given by (\ref{eq:opt_problem_simp}).
\end{lem}
\begin{IEEEproof}
First,  for all $i$, we show that $\bt_i^*\bt_i=1$ by induction. Since $\bt_1=\be_{1}$, we have $\bt_1^*\bt_1=1$. By construction, we have $\bt_{i-1}^{*} \be_{i} = 0$. Suppose that $\bt_{i-1}^*\bt_{i-1}=1$ for some $i.$ Then, from (\ref{eq:t_construc}) and (\ref{eq:ci}), we have
\beq
\bt_i^*\bt_i &=& \frac{a_{i-1}^2}{a_i^2}\bt_{i-1}^*\bt_{i-1} + \frac{c_{i}^2}{a_i^2}, \n \\
&=& \frac{a_{i-1}^2}{a_i^2} + \frac{a_i^2- a_{i-1}^2}{a_i^2}, \n \\
\label{eq:tinorm}
&=& 1.
\eeq

Next, for all $i$, we show that $\bt_i^*\bc=a_i$ by induction. Since $\bt_1=\be_{1}$, we have $\bt_1^*\bc=a_1$. Suppose that $\bt_{i-1}^*\bc=a_{i-1}$ for some $i.$ Then, from (\ref{eq:t_construc}) and (\ref{eq:ci}), we have
\beq
\bt_i^*\bc &=& \frac{a_{i-1}}{a_i}\bt_{i-1}^*\bc + \frac{c_{i}}{a_i} c_{i}, \n \\
&=& \frac{a_{i-1}^2}{a_i} + \frac{a_i^2- a_{i-1}^2}{a_i}, \n \\
\label{eq:ta_property}
&=& a_i.
\eeq

Last, for all $i\le j$, using lower-triangular property of $\bT^*$ and (\ref{eq:ta_property}), we show that the $(i,j)$-th entry of $\bT^*\bG$ is equal to $h_{i,j}$:
\beq
(\bT^*\bG)_{i,j} &=& \bt_i^*b_j[d_{1,j}c_1 \: d_{2,j}c_2 \: \ldots \: d_{K,j}c_K  ]^*, \n \\
&=& \bt_i^*[c_1 \: c_2 \: \ldots \: c_K  ]^*b_j, \quad \forall i\le j,\n \\
\label{eq:TGupptriang}
&=& a_ib_j,\quad \forall i\le j.
\eeq
With (\ref{eq:tinorm}) and (\ref{eq:TGupptriang}), the proof is complete.
\end{IEEEproof}

Next, we show that parameters $d_{i,j}$ (for $i > j$) exist such that (\ref{eq:opt_problem_simp}) evaluates to RHS of (\ref{eq:ach_sum_rate}). For this, we consider a lower-triangular matrix $\bV$ with unit diagonal entries.  Let $(i,j)$-th entry of $\bV$ be denoted by $v_{i,j}$. Define $\bF = \bI + \bG\bG^*$. Therefore, from (\ref{eq:g_construc}), the $(i,j)$-th entry of $\bV\bF$ is
\beq
(\bV\bF)_{i,j}&=&\sum_{m=1}^{i}\left(v_{i,m}\left(\delta_{m,j} + \sum_{n=1}^{K}g_{m,n}g_{j,n} \right)\right),\n \\
&=&\sum_{m=1}^{i}v_{i,m}\delta_{m,j} + \n \\
\label{eq:VFij}
&&c_j\sum_{n=1}^{K}\left(b_n^2d_{j,n}\left(\sum_{m=1}^{i}v_{i,m} c_m d_{m,n}
\right)\right).
\eeq

Now, suppose that, for all $i\ge2$ and $n\le i-1$, the parameters are such that
\beq
\label{eq:hypothesis}
\sum_{m=1}^{i}v_{i,m} c_m d_{m,n} = 0, \quad \forall i\ge 2, n\le i-1.
\eeq
Then, for all $i$ and $j\le i$, substituting (\ref{eq:hypothesis}) and $d_{i,j} = 1$ for any $i \le j$ in (\ref{eq:VFij}) , we obtain
\beq
\label{VFsimplified}
(\bV\bF)_{i,j}=v_{i,j}+ c_j\sum_{n=i}^{K}\left(b_n^2\left(\sum_{m=1}^{i}v_{i,m} c_m
\right)\right),\forall i, j\le i.
\eeq
For the set of values given by
\beq
\label{eq:vij}
v_{i,j} =\frac{-c_ic_j\sum_{n=i}^{K}b_n^2}{\left(\sum_{m=1}^{i-1}c_m^2\right)\left(\sum_{n=i}^{K}b_n^2\right)+1}, \forall j < i,
\eeq
from (\ref{VFsimplified}), we have $(\bV\bF)_{i,j}=0$ for all $j < i$ (i.e., $\bV\bF$ is upper-triangular) and
\beq
(\bV\bF)_{i,i}&=&1+ c_i\sum_{n=i}^{K}\left(b_n^2\left(\sum_{m=1}^{i-1}v_{i,m} c_m
+c_i\right)\right),\n\\
&=&1+ \frac{c_i^2\left(\sum_{n=i}^{K}b_n^2\right)}{\left(\sum_{m=1}^{i-1}c_m^2\right)\left(\sum_{n=i}^{K}b_n^2\right)+1},\n\\
\label{eq:VFii}
&=&1+ \frac{(a_i^2-a_{i-1}^2)\left(\sum_{j=i}^{K}b_j^2\right)}{a_{i-1}^2\left(\sum_{j=i}^{K}b_j^2\right)+1},\quad \forall i.
\eeq
Substituting (\ref{eq:vij}) in (\ref{eq:hypothesis}), we obtain
\beq
\label{eq:hypothesis_simp}
c_i\left(\frac{-\sum_{m=1}^{i-1}\left(c_m^2 d_{m,n}\right)\sum_{j=i}^{K}b_j^2}{\left(\sum_{m=1}^{i-1}c_m^2\right)\left(\sum_{j=i}^{K}b_j^2\right)+1}  +  d_{i,n}\right)= 0,
\eeq
for all $i\ge 2$ and $n\le i-1$. For any given $n$, it is clear that we can choose $d_{i,n}$ for all $i>n$, such that (\ref{eq:hypothesis_simp}) is satisfied for all $i>n$. This directly follows form the fact these are linear equations in $d_{i,n}$ with same number of variables as equations. Therefore, we have a construction that satisfies the assumption in (\ref{eq:hypothesis}).

Now, for the above construction, $\bV\bF$ is upper-triangular and $|\bV|=1$. Therefore, from (\ref{eq:VFii}), we have \beq
\frac{1}{2}\log|\bF| &=& \frac{1}{2}\log|\bV \bF|= \frac{1}{2}\log\prod_{i=1}^K (\bV\bF)_{i,i},\n \\
&=& \frac{1}{2}\sum_{i=1}^K\log\left(1+ \frac{(a_i^2-a_{i-1}^2)\left(\sum_{j=i}^{K}b_j^2\right)}{a_{i-1}^2\left(\sum_{j=i}^{K}b_j^2\right)+1}\right),\n
\eeq
which exactly matches the achievable sum-rate in (\ref{eq:ach_sum_rate}).
\subsection{Sum Capacity}

The above analysis establishes the sum capacity of the class of $K$-user Gaussian degraded interference channels. We summarize this result in the following theorem.
\begin{thm}
Consider any $K$-user Gaussian degraded interference channel with $\bH = \ba \bb^*$, where $\ba=[a_1 \: a_2\: 
\ldots \: a_K]^*$ and $\bb=[b_1 \: b_2 \: 
\ldots \: b_K]^*$. Let  $a_1^2 \le a_2^2 \le  
\ldots \le a_K^2$ and $a_0=0$. Then, the sum capacity of this channel is 
\[
C^{\text{IC}}_{\Sigma}(\sqrt{{P}/{N}}\bH)=\frac{1}{2}\sum_{i=1}^K \log\left(1+ \frac{(a_i^2-a_{i-1}^2)\left(\sum_{j=i}^{K}b_j^2\right)P}{a_{i-1}^2\left(\sum_{j=i}^{K}b_j^2\right)P+N}\right).
\]
\end{thm}

\begin{rem}
This class of channels have degree of freedom equal to $1$. The degree of freedom can be obtained in a straightforward manner as the $K$-th receiver can decode messages from all transmitters. However, this approach does not give the required tight outer bound on sum rate.
\end{rem}

\section{Conclusion}
\label{sec:conclude}
In this paper, we develop a family of outer bounds for the $K$-user Gaussian interference channel based on constructing multiple-antenna genie-MAC receivers. This formulation results in an optimization problem that may not be easy to solve in the general case. We subsequently show that this family of outer bounds determine the exact sum capacity of the class of degraded interference channels, and provide closed-form expression for the sum capacity of $K$-user Gaussian degraded interference channels.



\begin{thebibliography}{10}
\providecommand{\url}[1]{#1}
\csname url@samestyle\endcsname
\providecommand{\newblock}{\relax}
\providecommand{\bibinfo}[2]{#2}
\providecommand{\BIBentrySTDinterwordspacing}{\spaceskip=0pt\relax}
\providecommand{\BIBentryALTinterwordstretchfactor}{4}
\providecommand{\BIBentryALTinterwordspacing}{\spaceskip=\fontdimen2\font plus
\BIBentryALTinterwordstretchfactor\fontdimen3\font minus
  \fontdimen4\font\relax}
\providecommand{\BIBforeignlanguage}[2]{{%
\expandafter\ifx\csname l@#1\endcsname\relax
\typeout{** WARNING: IEEEtran.bst: No hyphenation pattern has been}%
\typeout{** loaded for the language `#1'. Using the pattern for}%
\typeout{** the default language instead.}%
\else
\language=\csname l@#1\endcsname
\fi
#2}}
\providecommand{\BIBdecl}{\relax}
\BIBdecl

\bibitem{Ahlswede71}
R.~Ahlswede, ``Multi-way communication channels,'' in \emph{Proc. 2nd Int.
  Symp. Information Theory}, 1971, pp. 103--135.

\bibitem{Han_Kobayashi}
T.~Han and K.~Kobayashi, ``A new achievable rate region for the interference
  channel,'' \emph{IEEE Trans. Inform. Theory}, vol.~27, no.~1, pp. 49--60,
  1981.

\bibitem{Sato}
H.~Sato, ``The capacity of the {G}aussian interference channel under strong
  interference (corresp.),'' \emph{IEEE Trans. Inform. Theory}, vol.~27, pp.
  786--788, 1981.

\bibitem{Shang}
X.~Shang, G.~Kramer, and B.~Chen, ``A new outer bound and the
  noisy-interference sum-rate capacity for {G}aussian interference channels,''
  \emph{IEEE Trans. Inform. Theory}, 2007.

\bibitem{Motahari}
A.~Motahari and A.~Khandani, ``Capacity bounds for the {G}aussian interference
  channel,'' \emph{IEEE Trans. Inform. Theory}, vol.~55, no.~2, pp. 620 --643,
  feb. 2009.

\bibitem{Annapureddy}
V.~Annapureddy and V.~Veeravalli, ``Gaussian interference networks: Sum
  capacity in the low interference regime,'' in \emph{IEEE Trans. Inform.
  Theory}, 6-11 2008, pp. 255 --259.

\bibitem{ETW}
R.~Etkin, D.~Tse, and H.~Wang, ``{G}aussian interference channel capacity to
  within one bit: the general case,'' \emph{Proc. IEEE International Symposium
  on Information Theory}, 2007.

\bibitem{Kramer00}
G.~Kramer, ``Outer bounds on the capacity of {G}aussian interference
  channels,'' \emph{IEEE Trans. Inform. Theory}, vol.~50, pp. 581--586, Mar.
  2004.

\bibitem{Cadambe-Jafar}
V.~R. Cadambe and S.~A. Jafar, ``Interference alignment and the degrees of
  freedom for the {K}-user interference channel,'' \emph{IEEE Transactions on
  Information Theory}, vol.~54, no.~8, pp. 3425--3441, Aug 2008.

\bibitem{Jafarian-Jose-Vishwanath-09}
A.~Jafarian, J.~Jose, and S.~Vishwanath, ``Algebraic lattice alignment for
  {K}-user interference channels,'' in \emph{Proc. Allerton Conference on
  Commun., Control and Computing}, Oct. 2009, pp. 88 --93.

\bibitem{Jafar}
V.~R. Cadambe and S.~A. Jafar, ``Multiple access outerbounds and the
  inseparability of parallel {Gaussian} interference channels,'' \emph{Proc.
  {IEEE} Globecom}, 2008.

\bibitem{EO}
R.~Etkin and E.~Ordentlich, ``On the degrees-of-freedom of the {K}-user
  {G}aussian interference channel,'' \emph{IEEE Trans. Inform. Theory}, 2009.

\bibitem{Sato_degraded}
H.~Sato, ``On degraded gaussian two-user channels (corresp.),'' \emph{IEEE
  Trans. Inform. Theory}, vol.~24, no.~5, pp. 637 -- 640, sep 1978.

\bibitem{Sason}
I.~Sason, ``On achievable rate regions for the gaussian interference channel,''
  \emph{IEEE Trans. Inform. Theory}, vol.~50, no.~6, pp. 1345 -- 1356, june
  2004.

\end{thebibliography}
\end{document}